\documentclass[prd,preprint,nofootinbib,superscriptaddress]{revtex4}
\usepackage[latin9]{inputenc}
\usepackage{amsmath}
\usepackage{amssymb}
\usepackage{color}
\usepackage{graphicx}
\usepackage{amsfonts}
\usepackage{latexsym}
\usepackage{epsfig}
\usepackage{hyperref}
\usepackage{subcaption}

\newcommand{\e}{\mathrm{e}}
\newcommand{\dee}{\mathrm{d}}

\makeatletter

\@ifundefined{textcolor}{}
{%
 \definecolor{BLACK}{gray}{0}
 \definecolor{WHITE}{gray}{1}
 \definecolor{RED}{rgb}{1,0,0}
 \definecolor{GREEN}{rgb}{0,1,0}
 \definecolor{BLUE}{rgb}{0,0,1}
 \definecolor{CYAN}{cmyk}{1,0,0,0}
 \definecolor{MAGENTA}{cmyk}{0,1,0,0}
 \definecolor{YELLOW}{cmyk}{0,0,1,0}
}

\makeatother

\begin{document}

\title{Electron mass variation from dark sector interactions and compatibility with cosmological observations}
\author{Kouki Hoshiya}
\email{k-hoshiya@particle.sci.hokudai.ac.jp}

\affiliation{Department of Physics, Hokkaido University, Sapporo 060-0810, Japan}
\author{Yo Toda}
\email{y-toda@particle.sci.hokudai.ac.jp}

\affiliation{Department of Physics, Hokkaido University, Sapporo 060-0810, Japan}

\begin{abstract}
We investigate the model where electrons and dark matter interact with dark energy through the rolling of a scalar field which comes from extra dimensional theories such as the braneworld theory and Brans-Dicke theory. In this model, dark energy couples to dark matter and electrons, which leads to larger values of the mass energies of dark matter and electrons in the early universe. We also fit our model to the cosmological data. By analyzing the data from Planck, baryon acoustic oscillation (BAO), light curves (Pantheon), and type-Ia supernovae (SH0ES), it can be seen that the Hubble tension is relieved in our model and the coupling parameter prefers a nonzero value with a significance of over $2\sigma$.

\end{abstract}

\maketitle
\preprint{EPHOU-22-005}

\vspace*{1cm}





\section{Introduction}

The $\Lambda$CDM model has been successful in 
explaining the properties and observations of our universe. However, there is a discrepancy between the value of
the Hubble constant reported from indirect measurements and the value from direct measurements.

Indirect measurements by Planck \cite{Aghanim:2018eyx} and the Atacama
Cosmology Telescope  \cite{ACT:2020gnv}, which observed cosmic microwave background
(CMB) anisotropies, give a value for the Hubble constant $H_0=67.36 \pm 0.54$km/s/Mpc,  $H_0=67.9 \pm 1.5$km/s/Mpc, respectively. An analysis \cite{DES:2017txv}, which is independent of CMB and which combined the Dark Energy Survey (DES), baryon acoustic oscillation (BAO), and big bang nucleosynthesis (BBN), reported the value of the Hubble constant as $H_0= 67.2^{+1.2}_{-1.0}$km/s/Mpc. 
On the other hand, local measurements of $H_0$ by Riess \cite{Riess:2019cxk, Riess:2020fzl}, 
which used Cepheids and supernovae, reported the value of Hubble constant as
 $H_0 = 74.03\pm1.42$km/s/Mpc(2019), $H_0 = 73.2\pm1.3$km/s/Mpc(2020), respectively.
Also, the H0LiCOW collaboration with lensed quasars \cite{Wong:2019kwg} reported as $H_0=73.3^{+1.8}_{-1.7}$km/s/Mpc and the observation using the Tip of the Red Giant Branch (TRGB) as distance ladders \cite{Freedman:2020dne} reported as $H_0=69.6\pm0.8(\mathrm{stat}) \pm 1.7(\mathrm{sys})$km/s/Mpc. Thus, the tension of the values of the Hubble constant between indirect measurements
and local measurements (Hubble tension) is a significant problem, 
although statistical errors of the Planck might cause the tension~\cite{Efstathiou:2013via,Freedman:2017yms,Rameez:2019wdt,Ivanov:2020mfr}.

In this paper, we explore the possibility of electron mass variation from dark sector interaction to solve this Hubble tension.
In our model, matter components couple to the scalar field $\phi$,
which is responsible for dark energy (DE), through the mass of the form like $m_0\e^{\beta \phi}$. Once the scalar field  evolves, due to the interaction,
energy of elementary particles and dark matter (DM) flows into that of DE. As a result, the masses of the elementary particles and the DM become lighter. 
Particularly, among the elementary particles, electron mass crucially contributes to the cosmological evolution.
Therefore, we investigate the coupling dependence of the CMB power spectrum and cosmological parameters.
Through the investigation, we conclude that this scenario relieves the Hubble tension through the electron mass reduction
as is described in the previous studies~\cite{Planck:2014ylh,Sekiguchi:2020teg,Sekiguchi:2020igz,Solomon:2022qqf,Hart:2017ndk,Fung:2021wbz}. 

Here, we should emphasize the worth of our model or the significance of adding DM-DE interaction.
Our model is inspired by cosmological models with extra dimensions such as  heterotic M-theory~\cite{Lukas:1998yy}, Brans-Dicke theory~\cite{Wetterich:1994bg}, and the Randall-Sundrum-I (RS I) model~\cite{Brax:2002nt,Rhodes:2003ev,Davis:2005au,Randall:1999ee} (review papers are~\cite{Kim:2003pc,Maartens:2010ar}).
In the five-dimensional effective theory of these models, DE interacts with not only elementary particles but also DM.
Therefore, we investigate the model which includes the dark sector interaction whose contribution is widely discussed in the previous works~\cite{Kumar:2016zpg, Gumjudpai:2005ry, Valiviita:2008iv, Costa:2013sva, He:2010im, Dutta:2010yu, Koyama:2009gd, Simpson:2010vh, Wands:2012vg, Wang:2013qy, Xu:2013jma, Li:2013bya, Ferreira:2014jhn, Ali-Haimoud:2015pwa, Garcia-Aspeitia:2016kak, CarrilloGonzalez:2017cll, Yang:2018euj, Goswami:2019zci, Mukhopadhyay:2019jla, Cheng:2019bkh, Jimenez:2021ybe,Yang:2019vni,DiValentino:2019jae}.

Our work is also motivated by a phenomenological motivation. There have been some works which studied models with interactions between DE and baryons~\cite{Vagnozzi:2019kvw, Jimenez:2020ysu} and they relieve the Hubble tension a little (the DE-baryon interaction is also discussed in the context of the direct detection of DE~\cite{Vagnozzi:2021quy,Ferlito:2022mok}). In this paper, we focus on electrons instead of the baryons as the matter which interacts with DE and explore the possibility to approach the Hubble tension problem.

This paper is organized as follows. In Sec. II, we present our model setting. In Sec. III, we will see the method of our analysis and datasets which we use.  In Sec. IV, we give our result and analysis. In the Sec. V, we summarize this paper.


\section{Modeling}

\subsection{Background evolution}

The model which we discuss is based on the Randall-Sundrum-I (RS I) model~\cite{Randall:1999ee}, in which there are two branes. It is known that this model implies the existence of two scalar fields, $\phi_1$ and $\phi_2$, in the low energy region. One of the two fields corresponds to a bulk scalar field, which can propagate in the bulk space between the two branes, while the other field is related to the physical distance between the two branes. These scalar fields couple to matter on the branes differently. In this paper, we will focus on one of them which can evolve in time, or in the evolution of the universe, which is denoted by $\phi$ from now on. We also focus on one of the two branes, the visible brane, for simplicity.  

Using this idea we will see a possibility that in addition to masses of dark matter (DM), masses of elementary particles (e.g. electrons) can be varied through the interaction with the bulk scalar field $\phi$. 
We assume, however,  that masses of baryons are varied little since the masses of the elementary 
particles which compose nucleons (and baryons) are generally much less than the masses of the nucleons.

With these ideas in mind, the action which we discuss has the form in the Einstein frame \cite{Rhodes:2003ev}
\begin{equation}
S = \int \dee^4 x \sqrt{-g} \left[ \frac{R}{2 \kappa^2} -\frac{1}{2}   g^{\mu\nu} (\partial_{\mu} \phi )(\partial_{\nu} \phi ) - V(\phi) \right] + S_{\mathrm{matter}} (\psi , A(\phi) g_{\mu\nu} )  , 
\end{equation}
where $S_{\mathrm{matter}}$ is the Lagrangian for matters on the visible brane, $R$ is the Ricci scalar, $\psi$ is the matter field on the brane. Also the quantity $A$ is written as 
\begin{equation}
   A = \mathrm{exp} \left( 2\beta \phi \right),  
\end{equation}
where $\beta$ is a negative constant. Henceforth, we assume the derivative of the potential $V$ to be negligible in order to compare our theory with the $\Lambda$CDM model.

Since we assume that the universe is homogeneous, isotropic, and flat, we have the line segment of the form
\begin{equation}
   \dee s^2 = a^2(\tau) (-\dee \tau^2 + \delta_{\mathrm{ij}} \dee x^{\mathrm{i}} \dee x^{\mathrm{j}} ) . 
\end{equation}

Then we obtain the field equations  
\begin{equation}\label{eq st 1}
\mathcal{H}^2= \frac{1}{3} a^2 \left( \rho_{\mathrm{total}} + \frac{1}{2 a^2} \dot{\phi}^2 + V(\phi) \right);
\end{equation}
\begin{equation}\label{eq st 2}
  \ddot{\phi} + 2\mathcal{H} \dot{\phi}  = - \sum_{(\mathrm{i})}\beta( \rho_{(\mathrm{i})} - 3 p_{(\mathrm{i})} ) a^2 ;  
\end{equation}
\begin{equation}\label{eq st 3}
   \dot{\rho}_{(\mathrm{i})} + 3 \mathcal{H} (\rho_{(\mathrm{i})} + p_{(\mathrm{i})}) = \beta (\rho_{(\mathrm{i})} - 3 p_{(\mathrm{i})}) \dot{\phi},  
\end{equation}
where the dot denotes the derivative with respect to the conformal time $\tau$ and $\mathcal{H}:= \dot{a}/a$. Note that $\rho_{(\mathrm{i})}$ runs over density of cold dark matter $\rho_{\mathrm{c}}$ and density of elementary particles $\rho_{\mathrm{ep}}$
\footnote{We may include density of radiations as well. However the energy-stress tensor of the radiations is traceless and the right-hand sides of the Eqs. \eqref{eq st 2} and \eqref{eq st 3} vanish, which are not interesting. }
, while the $\rho_{\mathrm{total}}$ contains all densities (including density of the baryons $\rho_{\mathrm{b}}$).  Hence we can rewrite the Eq. \eqref{eq st 3} as
\begin{equation}
    \dot{\rho}_{(\mathrm{i})} + 3 \mathcal{H} \rho_{(\mathrm{i})} = \beta \rho_{(\mathrm{i})} \dot{\phi} ,      
\end{equation}
where the subscript i can be either ``c'' or ``ep''. The solution to this equation has the form
\begin{equation}\label{sol of densities}
    \rho_{(\mathrm{i})} = \rho_{(\mathrm{i})0} a^{-3} \e^{\beta \phi},
\end{equation}
where $\rho_{(\mathrm{i})0}$ is a constant and this implies that we can express the variation of the masses as
\begin{equation}
    \frac{m_{(\mathrm{i})}}{m_{\mathrm{(i)0}}} \propto \sqrt{A} = \e^{\beta \phi} .  
\end{equation}
This formula shows us the explicit relation between the evolution of the bulk scalar field $\phi$ and the evolution of the masses of matters which interact with the scalar field $\phi$. 

\begin{figure}[h]
\hspace{-6.5mm}
\begin{subfigure}{.51\textwidth}
  \centering
  \includegraphics[width=.86\linewidth]{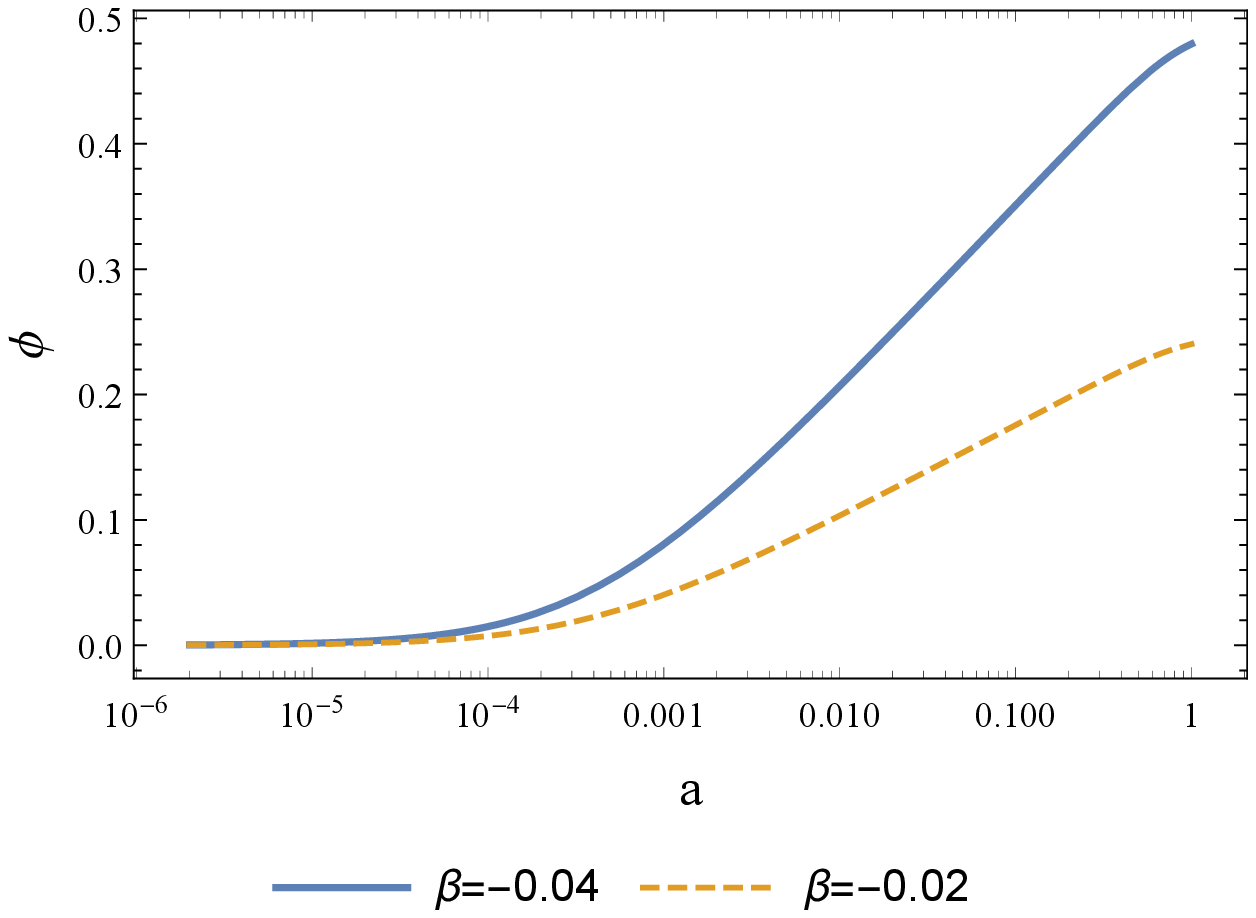}
  \caption{Evolution of the scalar field $\phi$}
  \label{fig:1a}
\end{subfigure}%
\begin{subfigure}{.51\textwidth}
  \centering
  \includegraphics[width=.9\linewidth]{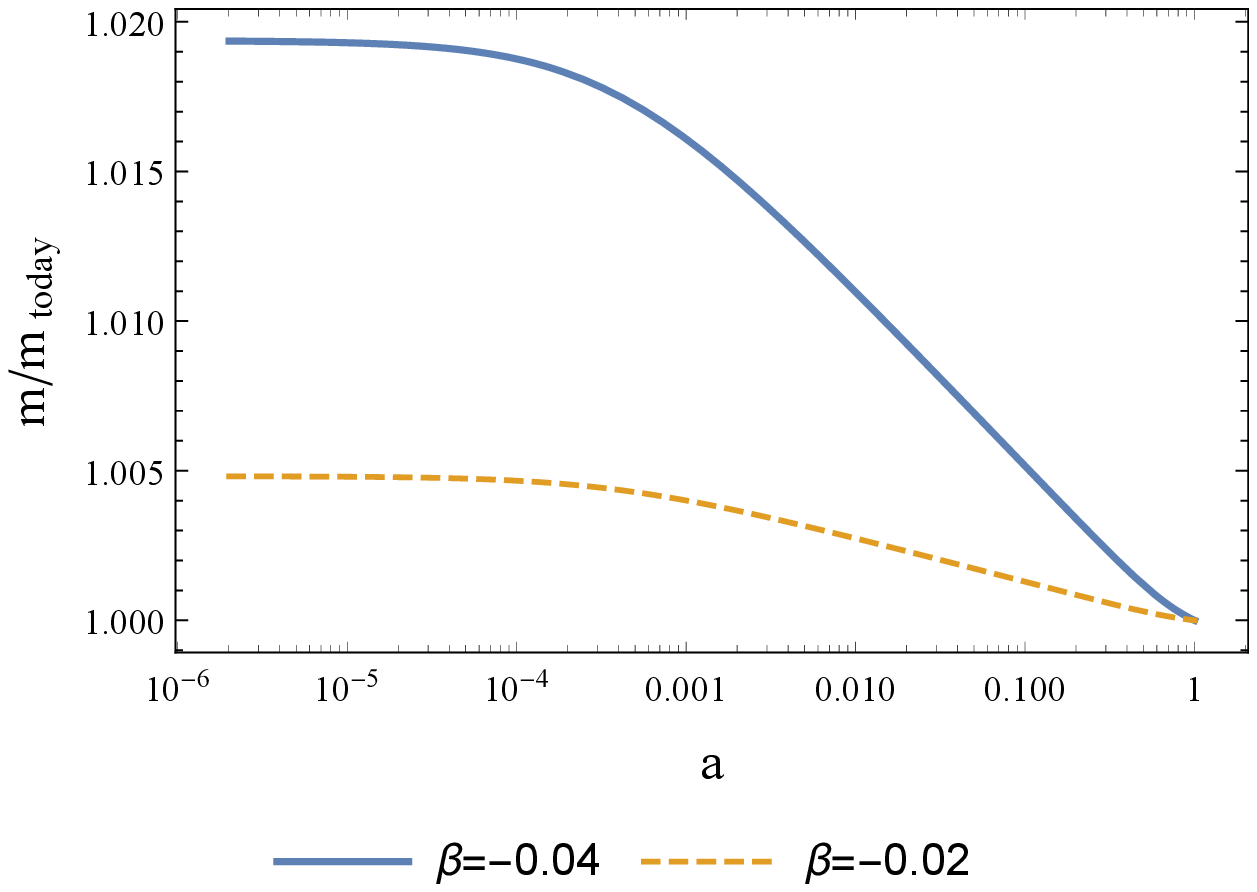}
  \caption{Evolution of the mass}
  \label{fig:1b}
\end{subfigure}
\vspace{-6mm}
\caption{We have set the initial value of $\phi$ as zero and set the ratio of the baryon density $\rho_{\mathrm{b}}$ over the all matter densities as $0.2$.}
\label{fig:1}
\end{figure}

On the other hand, the baryons do not interact with the scalar field $\phi$, since as we have mentioned at the beginning of this section, the mass of the baryons is assumed to be invariant and this means that the energy density of the baryons $\rho_{\mathrm{b}}$ is also invariant due to the fact that the baryons are nonrelativistic particles. Therefore, the counterpart of the Eq. \eqref{eq st 3} for them becomes 
\begin{equation}
    \dot{\rho}_{\mathrm{b}} + 3 \mathcal{H} \rho_{\mathrm{b}} = 0,  
\end{equation}
which leads to the solution
\begin{equation}\label{sol of baryon}
    \rho_{\mathrm{b}} = \rho_{\mathrm{b}0} a^{-3} ,
\end{equation}
where $\rho_{\mathrm{b}0}$ is an arbitrary constant. The numerical solutions to the Eqs. \eqref{eq st 1}, \eqref{eq st 2}, and \eqref{eq st 3} are given in Fig. \ref{fig:1}. Note that in matter dominant era, the scalar field $\phi$ evolves in logarithmic way while in the radiation dominant era, that behaves as almost constant.

The mass variation is also shown in Fig. \ref{fig:1}.
Note that the electron mass contributes to the energy levels of a hydrogen atom ($\varpropto m_{\mathrm{e}}$)
and Thomson scattering cross section $\sigma_{\mathrm{T}}$ ($\propto m_{\mathrm{e}}^{-2}$)~\cite{Planck:2014ylh,Sekiguchi:2020teg,Sekiguchi:2020igz,Solomon:2022qqf,Hart:2017ndk}.

Substituting the solutions \eqref{sol of densities} and \eqref{sol of baryon} to the rest equations \eqref{eq st 1} and \eqref{eq st 2} and exploiting the fact that $\rho_{\mathrm{c}}>\rho_{\mathrm{ep}}$, in the matter-dominated era we have
\begin{equation}\label{eq st 4}
   \mathcal{H}^2 = \frac{1}{3a} \left( \rho_{\mathrm{b}0} +  \rho_{\mathrm{c}0}  \e^{\beta \phi}  \right);   
\end{equation}
\begin{equation}\label{eq st 5}
  \ddot{\phi} + 2 \mathcal{H} \dot{\phi} = - \beta \, \rho_{\mathrm{c}0} a^{-1} \e^{\beta \phi}.     
\end{equation}
Here we set the today's energy ratio $\omega$ of matters and CDM as
\begin{equation}
  \omega = \frac{\rho_{\mathrm{c0}}}{\rho_{\mathrm{b0}}+\rho_{\mathrm{c0}}}.             
\end{equation}

As disccussed in the previous work~\cite{Rhodes:2003ev}, the solution of the scalar field $\phi$ to the system of equations has the form of $-2\beta \mathrm{ln} a$ in the matter-dominant and in the case that the matter consists of only dark matter. To take into account the contribution from the baryons, we have exploited a fitting formula for the numerical solution:
\begin{equation}\label{eq st 15}
   \phi = -2\beta \omega \left( \mathrm{ln}(a+a_{\mathrm{eq}}) - \mathrm{ln} (a_0 + a)  \right) + \phi_0.  
\end{equation}    
Note that this formula also fits to the numerical solutions in radiation-dominant and matter-DE-equality eras as well as in the matter-dominant era. Here $a_{\mathrm{eq}}$ and $a_{0}$ are the scale factors at the matter-radiation equality and matter-DE-equality, respectively, and $\phi_0$ is the initial value. Note that the $\mathrm{ln}(a+a_{\mathrm{eq}})$ behaves like $\mathrm{ln} \,a$ and $\mathrm{ln} \, a_{\mathrm{eq}}$ when $a > a_{\mathrm{eq}}$ and $a < a_{\mathrm{eq}}$, respectively.

\begin{figure}[h]
\hspace{-5mm}
\begin{subfigure}{.51\textwidth}
  \centering
  \includegraphics[width=.9\linewidth]{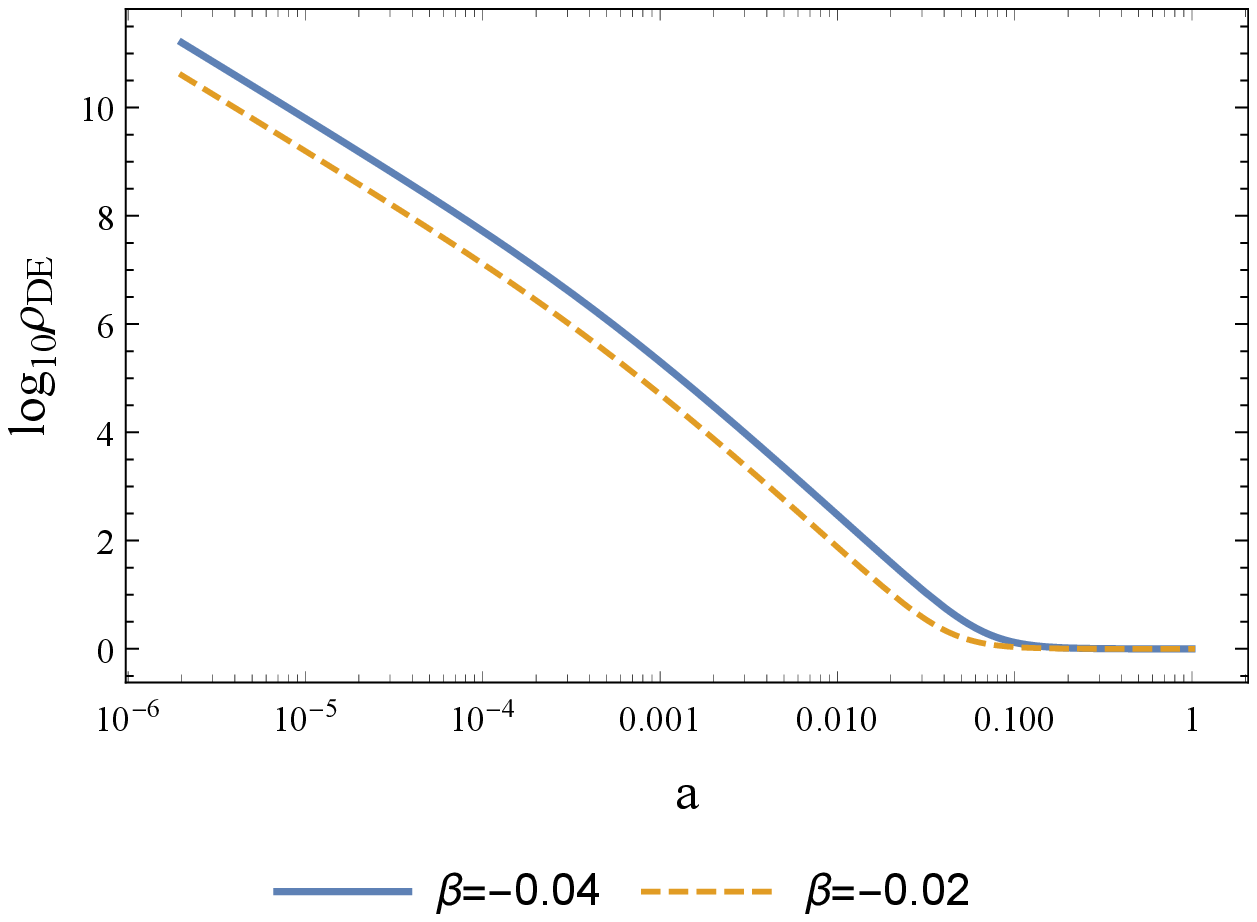}
  \caption{}
  \label{fig:2a}
\end{subfigure}%
\begin{subfigure}{.51\textwidth}
  \centering
  \includegraphics[width=.9\linewidth]{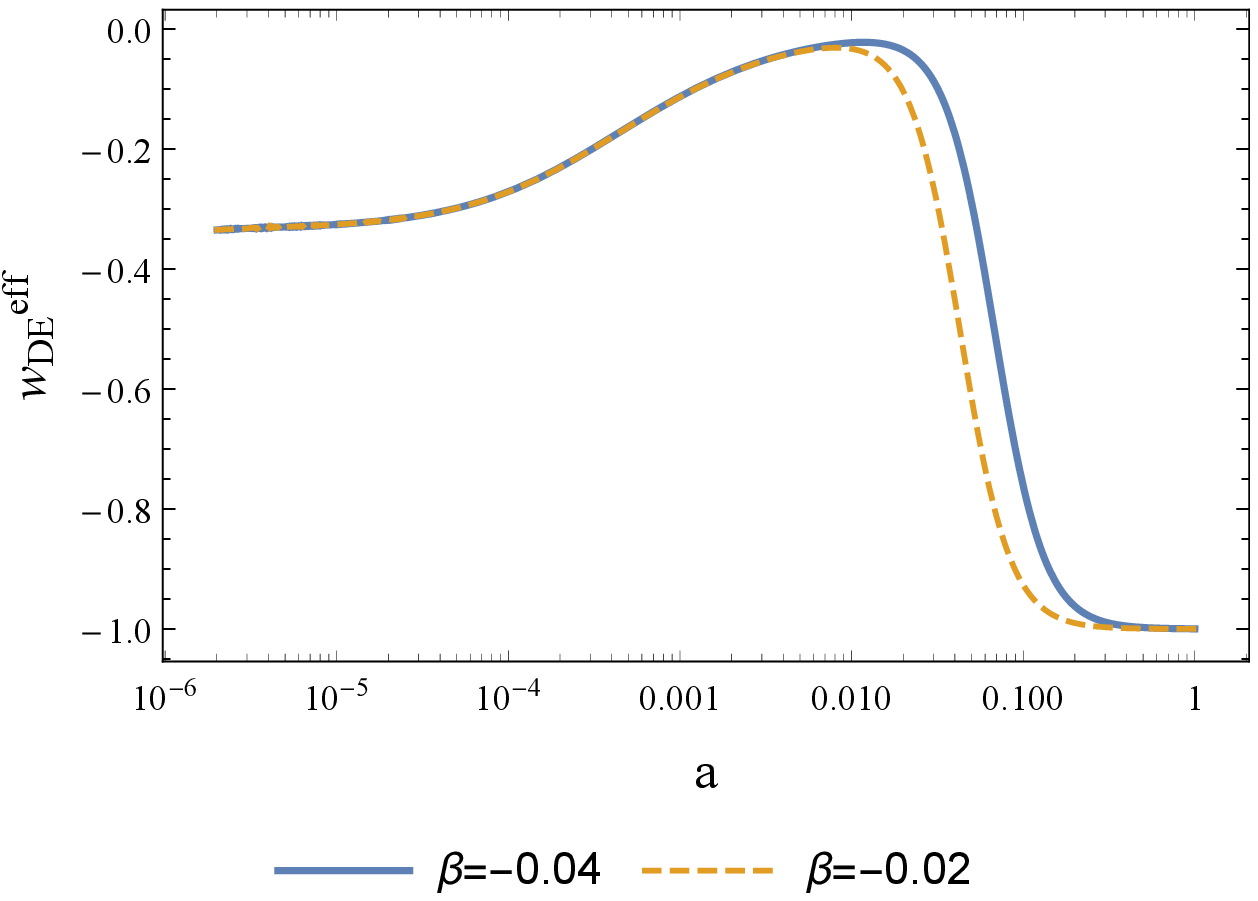}
  \caption{}
  \label{fig:2b}
\end{subfigure}
\vspace{-6mm}
\caption{Evolution of the density of DE $\rho_{\mathrm{DE}}$ (a) and density of the effective equation of state $w^{\mathrm{eff}}_{\mathrm{DE}}$ (b). We have set the value of $a$ as 1 at the present.}
\label{fig:2}
\end{figure}

To understand the evolution of the dark energy, we consider evolution of the effective equation of state $w^{\mathrm{eff}}_{\mathrm{DE}}$ (Fig. \ref{fig:2b}), which is defined as
\begin{equation}
    w^{\mathrm{eff}}_{\mathrm{DE}} = -1 - \frac{1}{3 \rho_{\mathrm{DE}}} \frac{\partial \rho_{\mathrm{DE}} }{\partial (\mathrm{ln} a)},   
\end{equation}
where $\rho_{\mathrm{DE}}= \dot{\phi}^2/(2a^2)  + V$ is the energy density of the dark energy.

We set  the critical point $a_{\mathrm{crit}}$ as the transition point when dark energy changes from kination dominant to potential dominant. Before the critical point $a_{\mathrm{crit}}$, since the potential $V$ has only less contribution, we can write down the $w^{\mathrm{eff}}_{\mathrm{DE}}$ as
\begin{equation}
     w^{\mathrm{eff}}_{\mathrm{DE}} \simeq -1 + \frac{2}{3} \left( 1 - \frac{\phi''}{\phi'}- \frac{\mathcal{H}'}{\mathcal{H}} \right)  , 
\end{equation}
where the prime denote the derivative with respect to $\mathrm{ln} a$. Using Eq. \eqref{eq st 15}, $\phi''/\phi'$ can be calculated as      
\begin{equation}
      \frac{\phi''}{\phi'} = \frac{-a^2 +a_0 a_{\mathrm{eq}} }{(a+a_0)(a+a_{\mathrm{eq}})}.  
\end{equation}
In the radiation dominant era, the value of $\phi''/\phi'$ becomes $1$, while the value of $\mathcal{H}'/\mathcal{H}$ becomes $-1$ so that the $w^{\mathrm{eff}}_{\mathrm{DE}}$ has the values asymptotically going to $-1/3$. In the matter dominant era, $\phi''/\phi'$ has the value 0 at a moment, while $\mathcal{H}'/\mathcal{H}$ has the value $-1/2$. As a result the maximum value of the $w^{\mathrm{eff}}_{\mathrm{DE}}$ close to 0 and it decreases with only small rate for a while as time goes back. 

After the critical point $a_{\mathrm{crit}}$, since the potential $V$ has significant contribution, the value of $w^{\mathrm{eff}}_{\mathrm{DE}}$ becomes $-1$.

\subsection{Perturbative equation}
The interaction changes the perturbative equation as well. In the synchronous gauge,  we have the line segment corresponding to the scalar perturbation of the metric
\begin{equation}
   \dee s^2 = a^2(\tau) \left\{ -\dee \tau^2 + \left[\left(1+\frac{h}{3}\right)\delta_{\mathrm{ij}}+\left(\partial_{\mathrm{i}}\partial_{\mathrm{j}} - \frac{1}{3} \delta_{\mathrm{ij}} \nabla^2\right) 6\eta\right]  \dee x^{\mathrm{i}} \dee x^{\mathrm{j}} \right\} . 
\end{equation}

Before the scalar field begin to roll, dark energy behaves as constant and
we set the dark energy perturbation $\delta_{\mathrm{de}}, \theta_{\mathrm{de}}=0$~\cite{Weller:2003hw}.
The equation of motion for the perturbed scalar field $\delta\phi$
is given in~\cite{Rhodes:2003ev}

\begin{equation}
\ddot{\delta\phi}+2\mathcal{H}\dot{\delta\phi}+\left(k^{2}+a^{2}\frac{\partial^{2}V}{\partial\phi}\right)\delta\phi+\frac{1}{2}\dot{h}\dot{\phi}=-\beta\rho_{\mathrm{c}}\delta_{\mathrm{c}}a^{2}
\end{equation}

With this equation and the relations $\rho_{\mathrm{de}}\delta_{\mathrm{de}}=a^{-2}\dot{\phi}\dot{\delta\phi}+V_{,\phi}\delta\phi$
and $(\rho_{\mathrm{de}}+p_{\mathrm{de}})v_{\mathrm{de}}=a^{-2}k\dot{\phi}\delta\phi$~\cite{Hu:1998kj},
where $\delta$ is the density fluctuation and $v$ is the velocity,
we modify the perturbative equation of dark energy as follows,

\begin{align}
\dot{\delta}_{\mathrm{de}} & =-3\mathcal{H}(1-w_{\mathrm{de}})\delta_{\mathrm{de}}-(1+w_{\mathrm{de}})kv_{\mathrm{de}}-(1+w_{\mathrm{de}})\frac{\dot{h}}{2}\nonumber \\
 & -9\mathcal{H}^{2}(1-c_{\mathrm{a}}^{2})(1+w_{\mathrm{de}})\frac{v_{\mathrm{de}}}{k}-\beta\delta_{\mathrm{de}}\frac{a^{2}\rho_{\mathrm{c}}}{\dot{\phi}}+\beta\frac{\rho_{\mathrm{c}}}{\rho_{\mathrm{de}}}\dot{\phi}(\delta_{\mathrm{de}}-\delta_{\mathrm{c}})\label{eq:15}
\end{align}

\begin{equation}
\dot{v}_{\mathrm{de}}=2\mathcal{H}v_{\mathrm{de}}+\frac{\delta_{\mathrm{de}}}{1+w_{\mathrm{de}}}k+\beta\dot{\phi}\frac{\rho_{\mathrm{c}}}{\rho_{\mathrm{de}}}v_{\mathrm{de}}\frac{c_{a}^{2}}{1+w_{\mathrm{de}}}
\end{equation}
where $c_{\mathrm{a}}^{2}\equiv\dot{p}_{\mathrm{de}}/\dot{\rho}_{\mathrm{de}}$
is the adiabatic sound speed and we substituted $1$ for the
sound speed $c_{\mathrm{s}}$. The nonperturbative part and the last
term of Eq. (\ref{eq:15}) are the same as the previous work~\cite{Costa:2013sva}
and the other terms are changed due to the different treatments
of the perturbations.

We also change the perturbative equation of DM as follows~\cite{Rhodes:2003ev},

\begin{equation}
\dot{\delta_{\mathrm{c}}}=-\left(kv_{\mathrm{c}}+\frac{\dot{h}}{2}\right)+\beta\dot{\delta\phi}
\end{equation}

\begin{equation}
\dot{v_{\mathrm{c}}}=-\mathcal{H}v_{\mathrm{c}}+\beta k\delta\phi-\beta\dot{\phi}v_{\mathrm{c}}
\end{equation}

We should note that even in the synchronous gauge, the velocity of
DM $v_{\mathrm{c}}$ is not zero and evolves due to the DM-DE interaction,
while we set $v_{\mathrm{c}}=0$ initially to reduce the degrees of
the freedom of the gauge. We do not change the perturbative equations
of the baryon, because baryons barely interact with the scalar field.

\section{Data and Analysis}

\label{sec:analysis}

We perform a Markov-Chain Monte Carlo (MCMC) analysis on the braneworld
model described in the previous section. We use the public MCMC code
\texttt{CosmoMC-planck2018}~\cite{Lewis:2002ah} and implement
the above braneworld scenarios in its equation file and recombination code \texttt{recfast} in \texttt{camb}~\cite{Lewis:1999bs}. In particular, \texttt{recfast} is modified to implement the effects of electron mass variation which we mentioned after eq.(11).

To implement the scale-factor dependence of dark energy, we use the approximation formula \eqref{eq st 15}.
Note that we have defined the parameter $\delta\equiv-2\beta^{2}\omega$ just for convenience and sample $\delta\in[-0.01,0.00]$ in addition to the standard cosmological parameters [$\omega_\mathrm{b}$, $\omega_\mathrm{c}$, $\theta_\mathrm{\star}$,
$A_\mathrm{s}$, $n_\mathrm{s}$, $\tau_{\mathrm{reio}}$]. (The amount of dark energy $\Omega_\mathrm{DE}$ (or the value of V) is derived from the standard parameters.) We set $\delta$ negative because we require $\beta$ to be real, which appears in the Eqs. (21)-(24).


We analyze models with referring the following cosmological observation
datasets: 
\begin{itemize}
\item The CMB measurements from Planck (2018)~\cite{Aghanim:2018eyx}: temperature and polarization likelihoods
for high $l$ ($l=30$ to $2508$ in TT and $l=30$ to $1997$ in
EE and TE) and low$l$ \texttt{Commander} and lowE \texttt{SimAll}
($l=2$ to $29$)  

\item Gravitation lensing from Planck~\cite{Aghanim:2018oex}

\item BAO
from 6dF~\cite{Beutler:2011hx}, DR7~\cite{Ross:2014qpa} and DR12~\cite{Alam:2016hwk}

\item the local measurement
of light curves from Pantheon~\cite{Scolnic:2017caz}

\item the local measurement of the Hubble
constant from the Hubble Space Telescope observation of Supernovae
and Cephied variables from SH0ES (R19)~\cite{Riess:2019cxk}

\end{itemize}

\section{Result and discussion}

\label{sec:results}

\begin{figure}[htbp]
\centering
\includegraphics[width=120mm]{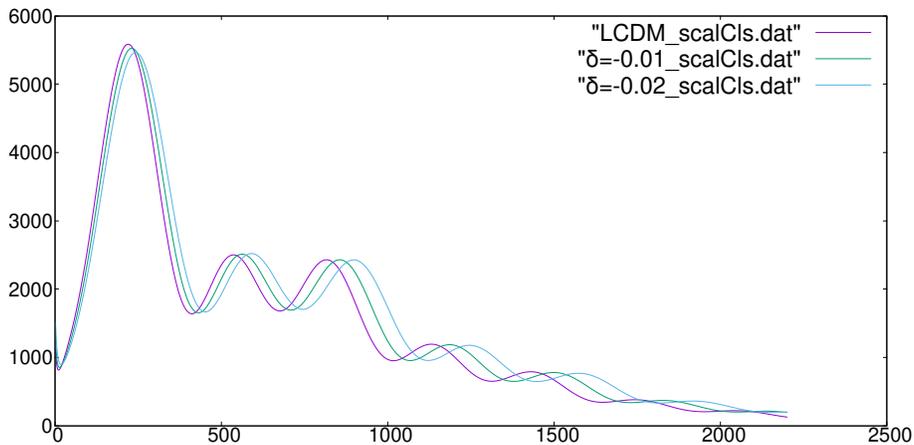}
\caption{CMB $TT$ angular power spectra for different values of $\delta$.}
\label{Power Spectrum}
\end{figure}

Figure \ref{Power Spectrum}\, shows the CMB power spectrum that we
compute using \texttt{camb}. We can find the two significant effects of our
model. First, the electron mass contributes to the recombination scale factor
$a_{*}$ through the energy levels of a hydrogen atom
and Thomson scattering cross section $\sigma_{\mathrm{T}}$~\cite{Planck:2014ylh,Sekiguchi:2020teg,Sekiguchi:2020igz,Solomon:2022qqf,Hart:2017ndk}.
Such contributions vary the redshift of the last scattering $z_{*}$ and sound horizon scale $r_{*}$, and shift the peak positions of the spectrum: lower
$\delta=-2\beta^{2}\omega$  (or greater $m_{\mathrm{e}}/m_{\mathrm{e}\,\mathrm{today}}$) leads to the higher redshift $z_{*}$ and shifts the
peak positions to higher multipole $l$. Second, the dark sector
interaction, where energy flows from DM into DE, leads
to the reduction in height of the first peak~\cite{Wang:2013qy}.
On the other hand, the second peak is a little amplified due to the earlier recombination which leads to the decrease of the Silk damping~\cite{Planck:2014ylh,Silk:1967kq}.

\begin{figure}[t]
\centering
\includegraphics[width=16cm]{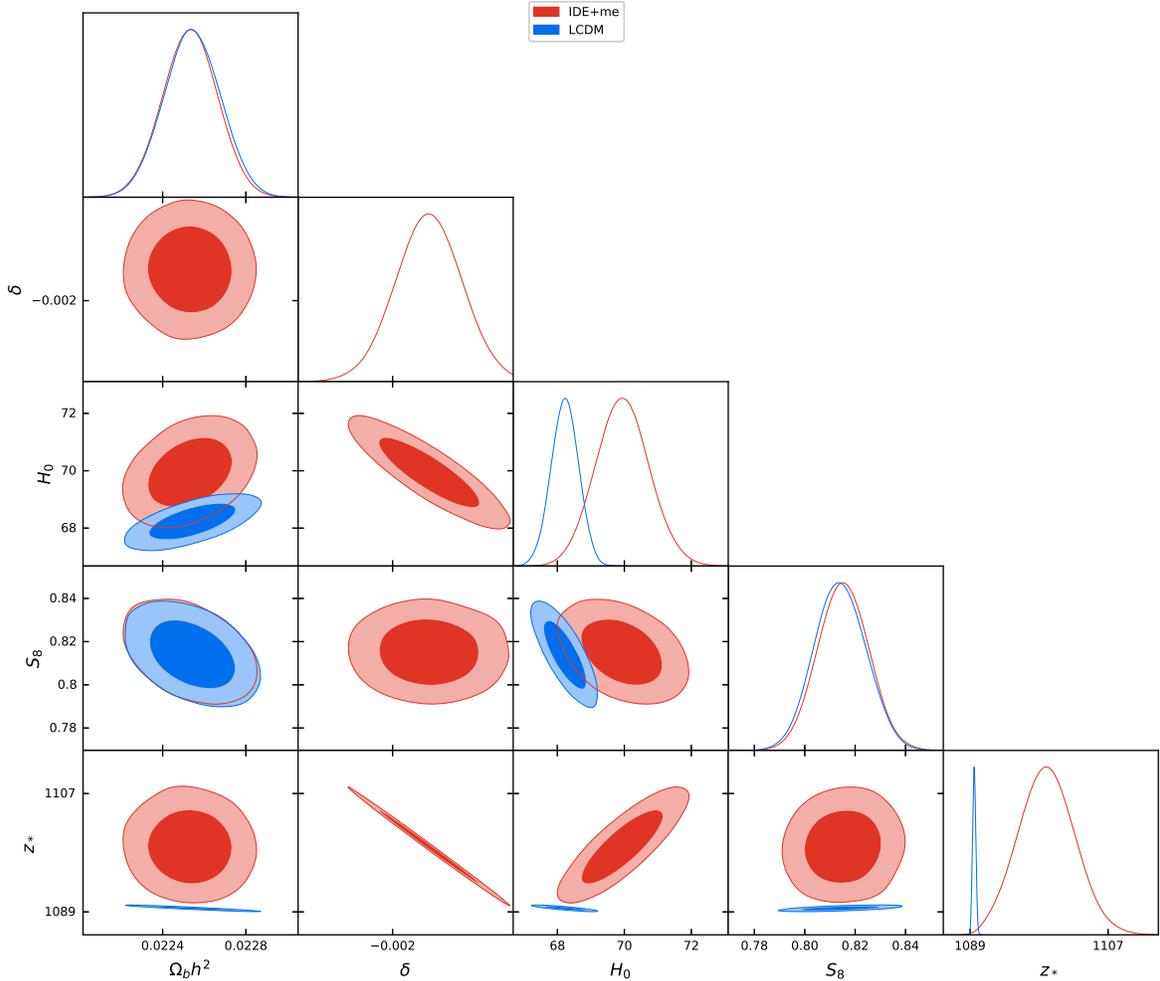} 
\caption{Posterior distributions of $\delta$, $H_0$, $S_8$, and $z_*$ on our model, which is called IDE-me. These posteriors have been derived for all datasets
(CMB+BAO+JLA+R19).}
\label{Fig:MC-1}
\end{figure}

\begin{figure}[t]
\centering
\includegraphics[width=16cm]{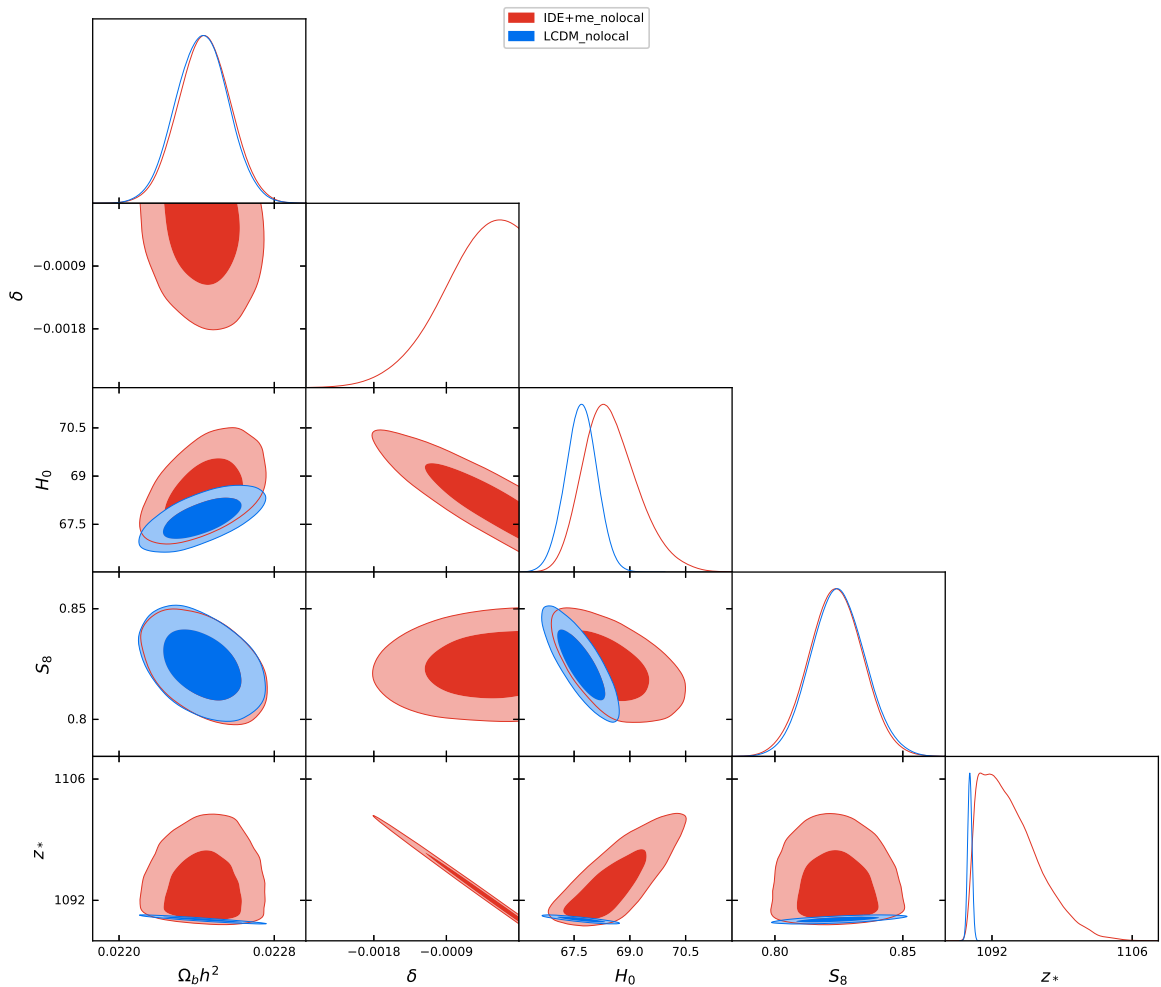} 
\caption{Posterior distributions of $\delta$, $H_0$, $S_8$, and $z_*$ on our model, which is called IDE-me. These posteriors have been derived for only distant datasets
(CMB+BAO).}
\label{Fig:MC-1-1}
\end{figure}

Figure \ref{Fig:MC-1} and Fig.\ref{Fig:MC-1-1} show the results of our
Monte Carlo analysis with the full dataset and with the only distant
datasets (CMB +BAO), respectively. 

From these figures, you can find that the Hubble tension
is relaxed in our model. This is due to the increment of the electron mass $m_{e}$, earlier recombination $z_{*}$, and shorter sound horizon $r_{*}$. 
We also show that the coupling parameter $\delta$ prefers a nonzero value with significance of over $2\sigma$ when we use the full datasets.

We obtain

\begin{align}
\delta(=-2\beta^{2}\omega)=-1.4_{-1.1}^{+1.1}\times10^{-3},\quad H_{0}=69.9_{-1.5}^{+1.6}\,\mathrm{km/s/Mpc},\\
(95\%,\mathrm{Planck}+\mathrm{BAO}+\mathrm{Pantheon}+\mathrm{SH0ES(R19)});\nonumber 
\end{align}
\begin{align}
\delta(=-2\beta^{2}\omega)>-1.4\times10^{-3},\quad H_{0}=68.5_{-1.3}^{+1.5}\,\mathrm{km/s/Mpc},\\
(95\%,\mathrm{Planck}+\mathrm{BAO}).\nonumber 
\end{align}

To compare the previous study of DE-baryon interaction~\cite{Jimenez:2020ysu}, we quote the result for $H_0=67.65_{-1.51}^{+1.52}$ (95\% Planck 2018 + CMB lensing + BAO + JLA + CFHTLensS + Planck SZ) and we conclude that our model relieves the Hubble tension more than the previous study. Eq.(26) reads that 2.4$\sigma$ tension remains between Planck+BAO and SH0ES in our model. However, the value of this tension is competitive with other 1-parameter solutions of the Hubble tension. (The tension values of other solutions are summarized in~\cite{Schoneberg:2021qvd}.)

We find another significant result that baryon fraction $\Omega_{\mathrm{b}}h^{2}$
does not increase, while in the simple electron mass varying model,
$\Omega_{\mathrm{b}}h^{2}$ increases. This means that our model does not
increase the baryon density discrepancy between the big-bang nucleosynthesis (BBN) analysis and
CMB measurements. This discrepancy has appeared by focusing on the
correlation between the baryon density and the deuterium abundance $D/H$
synthesized during the BBN~\cite{Cooke:2013cba, Cooke:2017cwo}. In the $\Lambda$CDM
model, using \texttt{PRIMAT}~\cite{Pitrou:2018cgg}, it is reported that this tension is $1.7\sigma$~\cite{Aghanim:2018eyx} and $1.8\sigma$~\cite{Pitrou:2020etk}
and it is also reported that in some Hubble tension solutions, the
increment of $\Omega_{\mathrm{b}}h^{2}$ makes this tension more
severe~\cite{Seto:2021xua}. However, our model does not.
This result supports the idea that we introduce the DM-DE interaction.
As we explained, DM-DE interaction suppresses the amplitude of the first peak of the CMB power spectrum, while the amplitude of the second peak is amplified due to the increment of the electron mass. Therefore, the increment of $\Omega_{\mathrm{b}}h^{2}$ is disfavored, which enlarges the difference between the first and second peaks.

As for the $S_{8}$ tension, our model does not
relieve the tension. We find that our model does not change the S8 and obtain
\begin{equation}
S_8=0.815^{+0.020}_{-0.019},~~\\
(68\%,\mathrm{Planck}+\mathrm{BAO}+\mathrm{Pantheon}+\mathrm{SH0ES(R19)}).
\end{equation}
Therefore, the tension remains with the Kilo-Degree Survey (KiDS)~\cite{Hildebrandt:2018yau} and Dark Energy Survey (DES)~\cite{DES:2017myr}, which give $S_{8}=0.737_{-0.037}^{+0.040}$ and $S_{8}=0.773_{-0.020}^{+0.026}$, respectively.

\begin{table}[h]
\[
\begin{tabular}{l||c|c}
Parameter &  ~\ensuremath{\Lambda}CDM~ & ~our\:model~\\
\hline \hline \ensuremath{\delta=-2\beta^{2}\omega}  &  0  &  -0.00104   \\
 \ensuremath{H_{0}} [km/s/Mpc]  &  68.17  &  69.54 \\
\hline   \ensuremath{\chi_{{\rm CMB}\:{\rm high}l}^{2}}  &  2346.31  & 2345.61 \\
 \ensuremath{\chi_{{\rm CMB}\:{\rm low}l}^{2}}  &  22.62  & 23.293  \\
 \ensuremath{\chi_{{\rm CMB}\:{\rm low}E}^{2}}  &  398.180  & 398.760 \\
 \ensuremath{\chi_{{\rm CMB}\:{\rm lensing}}^{2}}  &  8.595  & 8.852 \\
 \ensuremath{\chi_{{\rm H074p03}}^{2}}  &  16.983  &  9.980 \\
 \ensuremath{\chi_{{\rm JLA}}^{2}}  &  1034.80  &  1034.77  \\
 \ensuremath{\chi_{{\rm prior}}^{2}}  &  1.795  &  2.105 \\
 \ensuremath{\chi_{{\rm BAO}}^{2}}  &  5.200  & 6.386  \\
\hline  \ensuremath{\chi_{{\rm todal}}^{2}}  &  3834.47 & 3829.75 \\ 
\end{tabular} 
\]
\caption{The best-fit values of $\delta$, $H_0$, and $\chi^{2}$ for $\Lambda$CDM model and our model.}
\label{chi2}
\end{table}

The best-fit values of $\delta$, $H_0$, and $\chi^{2}$ from our model and $\Lambda$CDM model are summarized
in Tab.\ref{chi2}. The total $\chi^2$ of our model is less than the $\Lambda$CDM model.
The reduction in the value of $\chi_{{\rm total}}^{2}$ is mostly due to the improved fit of SH0ES measurement.
In addition, the slightly improved fit of CMB high $l$ spectra also decreases the value of $\chi_{{\rm total}}^{2}$.
However, the value of $\chi_{{\rm BAO}}^{2}$ increases in our model, which results from the modification of the low$z$ cosmology through the DM-DE interaction.
Therefore, we conclude that our model, which includes DE-DM \& DE-electron interaction,  is promising, although we should take care of BAO.

\section{Summary}

In this paper, we have studied the electron mass variation caused by dark sector interactions and researched the compatibility to the cosmological observations.
To sum up our model, the interaction rolls the scalar field $\phi$ of DE which couples to DM and elementary particles.
Then this leads to a phenomenon that the energy of DM and elementary particles (especially electrons, in this paper) transforms into that of DE. 
As a result, in the era of CMB, the masses of electrons and DM become larger than today.

We have performed the Monte Carlo analysis on our model with cosmological data.
As Fig.\ref{Fig:MC-1} and Fig.\ref{Fig:MC-1-1} show, our model prefers a larger Hubble constant.
Even the analysis with the only distant datasets gives the upper limit (95\%) of the Hubble constant as $70\mathrm{km/s/Mpc}$.
Although this value does not reach the value which is measured by the SH0ES measurement,
TRGB measurement is comparable with our model and we conclude that our model relives the Hubble tension.

The Table \ref{chi2} summarizes the $\chi^2$ of the each measurement.
This table shows that the total $\chi^2$ of our model is reduced by about 5 from $\Lambda$CDM model due to the improvement of the Hubble constant (SH0ES) and the slight improvement of the high $l$ CMB fit.

In our model, the scalar field of DE is rolled by the DE-DM interaction, and the electron mass is varied due to the coupling between electrons and DE. Of course, the rolling can be caused by the potential of the scalar field. For future work, such a model of DE-electron coupling is also worth considering.

\section*{Acknowledgments}
We would like to thank Dr. O. Seto for very useful advice and comments. This work was supported by JST SPRING, Grant No. JPMJSP2119.


\end{document}